\documentclass[]{article}
\usepackage[utf8]{inputenc}
\usepackage{geometry}
\usepackage[UKenglish]{babel}
\usepackage{amsmath, amsthm, amsfonts, authblk}
\usepackage[numbers, square]{natbib}
\usepackage{url}
\usepackage{graphicx}
\usepackage{lineno}
\usepackage{setspace}
\usepackage{todonotes}
\usepackage{subfigure}
\usepackage{multicol}
\usepackage[most]{tcolorbox}
\usepackage[labelfont=bf, textfont={small,sl}]{caption}

\begin{document}

\title{Characterising complex healthcare systems using network science: The small world of emergency surgery}

\author[1,*]{Katharina Kohler}
\author[1]{Ari Ercole}

\affil[1]{University Division of Anaesthesia, University of Cambridge, CB2 0QQ, UK}

\affil[*]{corresponding author: kk371@cam.ac.uk}

\date{}

\maketitle

\newcommand{\ari}[1]{\textbf{\color{cyan}[\textsc{Ari:} #1]}}
\newcommand{\kat}[1]{\textbf{\color{green}[\textsc{Kat:} #1]}}

\begin{abstract}
Hospitals are complex systems and optimising their function is critical to the provision of high quality, cost effective healthcare. Nevertheless, metrics of performance have to date focused on the performance of individual elements rather than the system as a whole. Manipulation of individual elements of a complex system without an integrative understanding of its function is undesirable and may lead to counter-intuitive outcomes and a holistic metric of hospital function might help design more efficient services. We aimed to characterise the system of peri-operative care for emergency surgical admissions in our tertiary care hospital using network analysis. We used retrospective electronic health record data to construct a weighted directional network of the system. For this we selected all unplanned admissions during a 3.5 year period involving a surgical intervention during the inpatient stay and obtained a set of 16,500 individual inpatient episodes.  We then constructed and analysed the structure of this network using established methods from network science such as degree distribution, betweenness centrality and small-world characteristics. The analysis showed the service to be a complex system with scale-free, small-world network properties. This finding has implications for the structure and resilience of the service as such networks, whilst being robust in general, may be vulnerable to outages at specific key nodes. We also identified such potential hubs and bottlenecks in the system based on a variety of network measures. It is hoped that such a holistic, system-wide description of a hospital service may provide better metrics for hospital strain and serve to help planners engineer systems that are as robust as possible to external shocks.

\end{abstract}

\flushbottom
\maketitle

\section*{Introduction}
Emergency surgical admissions make a significant contribution to pressures on hospitals and health services in general. This constitute an unpredictable high-resource patient group as their admissions are often characterised by intensive interventions and input from multiple teams. Such admissions are increasing: In October 2018 overall emergency admissions to hospital increased by 2\% compared to the previous year \cite{NHSEngland2018}. Admission pressure can only be accommodated if inpatient flow and discharges are as efficient and timely as possible and this may have knock-on effects all the way back to A\&E performance \cite{Keogh2018}. Increased emergency demand on a hospital affects elective services due to cancelled operations and increases in waiting times \cite{Keogh2018,Wong2018a}.

Service improvement projects and metrics typically focus either on single point interventions, specific periods within the patient journey or a defined subset of patients. Patient flow through hospitals has been investigated in various settings \cite{Martinez2018, Khanna2013, Khanna2016}, often in order to improve throughput in a specific setting such as the ED \cite{Afilal2016, Khanna2017}. However, hospitals and the multi-team services within them are considered `complex systems' \cite{Ebadi2019a}. Such highly interconnected and inter-dependent systems may respond counter-intuitively to interventions targeting individual modules and a system-wide, holistic description is lacking.

Network science, where the components of a system are represented by nodes and the connections by edges linking them, provides a method to investigate such systems and has been widely applied other disciplines \cite{Barabasi2016, Havlin2012, Newman2003, Eubank2004, Albert1999}. High-fidelity patient movement data is increasingly available in machine readable form making it available for modelling pathways in this way.

Recent papers used these methods in the medical services field to assess the structure of referral systems \cite{An2018}, surgical care teams \cite{Ebadi2019a} and patterns of acute hospital admissions \cite{Bean2017}. We wanted to assess whether we could apply network methodology to patient pathways throughout the hospital, specifically emergency surgical admissions. We attempted to construct a network representation of emergency surgical movements from patient movement data from our comprehensive Electronic Hospital Record (EHR) to give us a fuller understanding of the real architecture of the system of emergency surgical admissions to allow us to identify bottlenecks and vulnerabilities in the service structure.

\section*{Methods}

\subsection*{Data}
We used fully anonymised routinely collected location data from the EHR data at Cambridge University Hospitals NHS Foundation Trust, a tertiary care hospital. The hospital serves as a district general hospital to its local population and as a referral centre for specific specialties such as major trauma, neurosurgery and paediatric surgery. All major surgical specialties except for cardio-thoracic surgery are represented. We selected all unplanned admissions between January 2015 and July 2018 where the inpatient stay included a surgical procedure. This selection resulted in a dataset of more than 16,500 individual admissions of both adult and paediatric patients. Amongst the adult patients, the most common admitting surgical specialties were trauma (23.5\%), general surgery (13.7\%) and neurosurgery with (9.3\%). In addition to these surgical admitting specialties a significant proportion ($\sim21\%$ of patients) were admitted under medical specialties but subsequently required a surgical procedure during their stay.

Patient movement data including dates and times for each inpatient journey was collated from admission until discharge. We also included location data from inpatient investigations such as CT scans and inpatient visits to other departments. 

We used locations and transfers to develop our network model as each transfer represents a use of resources and is thought to be based on a medically necessary decision - e.g. a transfer to CT scan and back to the ward is related to the medical need for a scan and uses resources such as porters, ward staff and clinical staff to facilitate the event.

All data collected were extracted in a fully de-identified manner and stored securely. Under UK regulations, research ethics approval is not required for the re-use of anonymous routinely collected data for research. However our project was approved by local institutional review.

\subsection*{Model}
\label{subsec:model}

\subsubsection*{Network terminology and variables} \label{network}
A comprehensive mathematical description of network science concepts has been given by Barabasi \cite{Barabasi2016} and we include a short introduction to the concepts used in this paper. A network is formed of interconnected nodes and the number of connections a node has is called the \textit{degree} of the node. For directed networks, in-degree and out-degree for a specific node can also be calculated by counting the connections in or out. Connections between the nodes are then represented by their \textit{strength} or degree of a node weighted by the size of the edges. 

From this the degree probability distribution for all nodes can be constructed which contains information summarizing the underlying fundamental structure of the network. In a network with random node connections the degree distribution is a Poisson distribution, however in other types of networks other distributions such as log-normal or power law distributions with $p_{k}\sim  k^{-\gamma}$ can be fitted. Networks with a power-law distribution in the high degree tail arise when new nodes are preferentially attached to existing nodes with higher degree are known as `scale-free' \cite{Barabasi2003a}. Scale-free networks have a robust architecture as they are resistant to random node failure. This is due to the fact that most of the nodes have small degrees, so when a randomly selected node fails or is taken out of the network the impact is minimal. However, the presence of hubs (with disproportionately high degrees) makes these networks vulnerable to targeted insults where a hub is compromised \cite{Crucitti2004, Ebadi2019a}. 

It is then also possible to calculate a number of network properties including local and global \textit{clustering coefficients} (the extent to which node neighbours are linked), \textit{betweeness centrality} (identifying nodes with many shortest paths that are important to the overall functioning of the system), \textit{flow hierarchy} (the extent to which flow is directed),\textit{reciprocity} (the fraction of transfers that exist in both directions between two nodes) and \textit{assortativity} (the tendency of nodes to be connected to nodes of similar degree in our case).

Furthermore, the combination of clustering coefficient and shortest path length can then be used to assess the network for \textit{small-world} properties. Small-world networks are a type of network with specific properties such as high degree of clustering and clique formation, short average path between the nodes and an abundance of hubs, nodes with many connections\cite{Watts1998, Humphries2008, Telesford2011}. The measures $\sigma = \frac{\frac{C_{av}}{C_{rand}}}{\frac{L_{av}}{L_{rand}}}$ and $\omega = \frac{L_{r}}{Lav} - \frac{C_{av}}{C_{lat}}$ measure how close a network is to the small world ideal. The two characteristic measures are therefore a short path length and a high clustering coefficient. These properties are thought to facilitate easy and efficient flow of information and team work \cite{Ebadi2019a, Cowan2004} within the system. 

\subsubsection*{Network construction}
Patient locations were assigned to nodes and patient transfers to edges. We used the time-stamps to reconstruct each patient's journey. The data extraction from the EHR, data cleaning and construction of the network model was performed using python, networkx \cite{Hagberg2008}, igraph \cite{Csardi2006} and cytoscape \cite{shannon2003}. Data acquisition and handling was approved locally both via the trust QI (quality improvement) department and the university data request system. No further ethical approval was needed at no patients were directly contacted.

 We constructed both unweighted and weighted networks, where the weighting was the frequency of the transfer reflecting the traffic between two nodes. Since some specific wards could change their designation (e.g. from one specialty to another) or be re-purposed or closed without changing the underlying service, we subsequently created a `categorised' network by replaced agglomerating physical locations into care categories (E.g.  \textquotedblleft{acute medical ward}\textquotedblright).

\subsection*{Analysis}
We calculated the network characteristics such as degree distribution, strength distribution, flow hierarchy, global clustering and small-world properties described in the appendix.  Similarly, we investigated the properties of individual nodes with the measures described above, weighted and unweighted degree, clustering coefficients and betweenness centrality.

\section*{Results}
\subsection*{Overall network and degree distribution}\label{degreedist}

\begin{figure}[ht]
\centering
\includegraphics[width=0.8\textwidth]{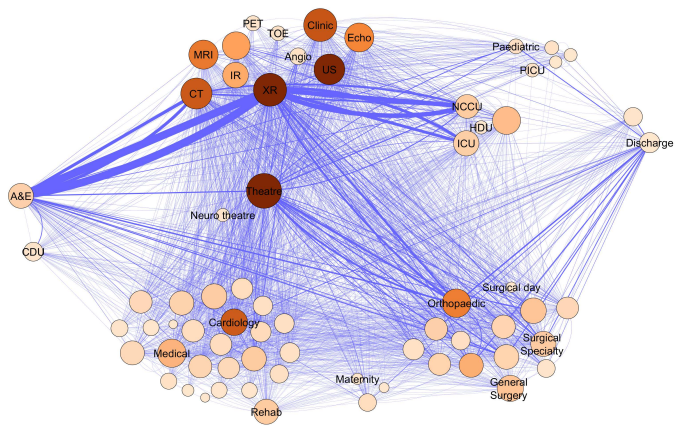}
\caption{\textbf{Non-categorised network of emergency surgical admissions} The network of transfers shown grouped by clinical categories of care. The nodes are colored by betweenness centrality - higher betweenness centrality is shown in deeper color- and sized by overall degree. The nodes were arranged to represent a patient journey from A\&E admission (on the left) to discharge (on the right) and the nodes were grouped together to show the different types of locations such as medical wards or high dependency areas. Some of the nodes have been labelled to show the grouping with the medical wards at the bottom left, the surgical wards bottom right, investigations at the top, critical care areas and theatres in the middle and the paediatric services at the top right. Most nodes are left unlabelled for clarity with some indicative labels in each group. The ward abbreviations are explained in the appendix.}
\label{fig:allstructured_network}
\end{figure}

The total cleaned dataset consisted of more than 16500 individual admissions with 230,000 transfers. The 'non-categorised' network of all transfers of patients admitted via A\&E is shown in Figure \ref{fig:allstructured_network}. The wards and services are grouped together by type of area to better illustrate the variety of services involved in the care of these patients. Descriptions of the abbreviations used can be found in the appendix. This overall network representation of our emergency surgical care network shows dense connectivity between the different wards and hospital services. Areas that are particularly densely connected (E.g. theatres or radiology)  most often have high overall network importance based on betweenness centrality. However, there are a few areas (E.g. medical wards) which are well connected but less structurally important. 

\begin{figure}[ht]
\centering
\includegraphics[width=0.8\textwidth]{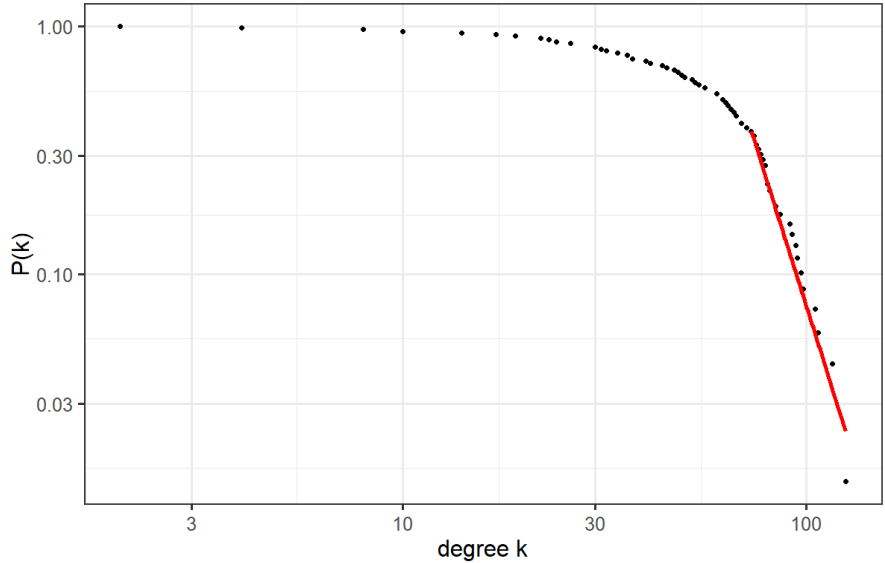}
\caption{\textbf{Degree distribution}. The degree distribution (as log-log plot) for our network of wards. The distributions shows a power-law behaviour at the right hand tail of the distribution. The power-law fit (obtained with the package poweRlaw \cite{powerlaw}) is shown in red with a $\gamma$ of 6.1.}
\label{fig:power_law}
\end{figure}

Figure \ref{fig:power_law} shows the degree distribution for our network. We used the methodology outlined in Clauset et al \cite{Clauset2009} to investigate the best fit to the tail of the distribution. The power-law fit generated using a bootstrap method \cite{Efron1979} to determine the most representative fit to the tail of the distribution $P(k)=k^{-\gamma}$ is superimposed in red. We found $\gamma =6.18$ (95\% CI 6.14 - 6.26) with a $p$-value for the null hypothesis of 0.46, which means that a power-law is an appropriate fit. 

\begin{figure}[ht]
\centering
\includegraphics[width=0.8\textwidth]{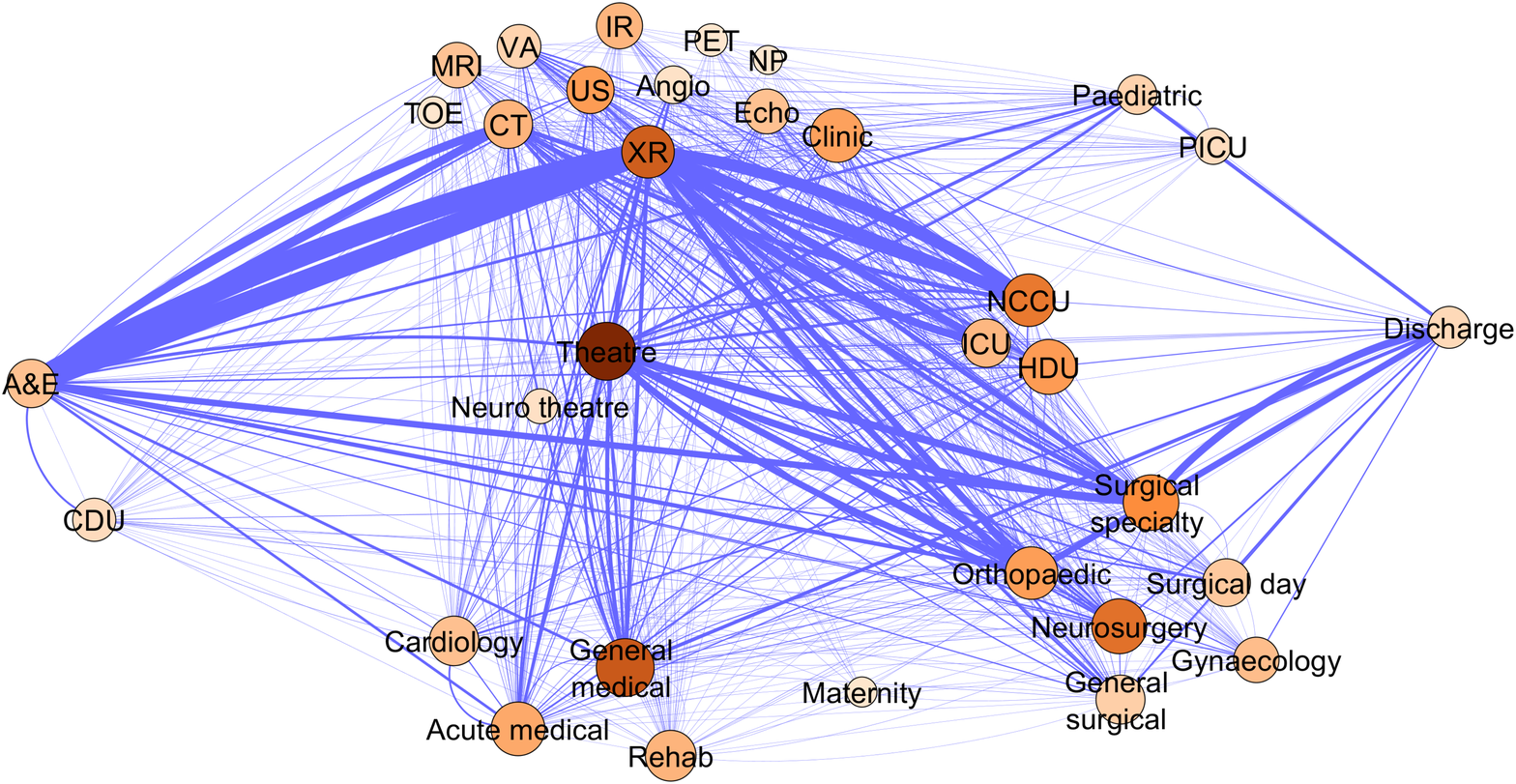}
\caption{\textbf{Categorised network} The network of transfers shown grouped by area of care. Here the nodes are categories rather than physical locations and as in Figure \ref{fig:allstructured_network} colored by betweenness centrality - higher betweenness centrality is shown in deeper color- and sized by overall degree. The ward abbreviations are contained in the appendix}
\label{fig:category_network}
\end{figure}

The categorised network is shown in Figure \ref{fig:category_network}. Some of strongest connections are unsurprisingly between ED and radiology or CT scan - illustrating the need for initial investigations upon a patient's admission to hospital. Other strong connections are more surprising: The presence of a connection from ED to the general medical wards and one from these wards to discharge shows that our patient cohort attended the medical wards both pre- and post-operatively. The betweenness centrality measure shows the importance of theatres, general medical and neurosurgical wards to the overall function of the system.

We found both our non-categorised and categorise networks had high reciprocity (with $r \sim 0.9$ and $r \sim 0.8$ respectively). This means that most paths exist in both directions and that transfers from one area to another are usually countered by transfers in the opposite direction-- not necessarily the same patient. This balance reflects the operational need to fill all beds and spaces to allow for the high demand within the service.

\subsection*{Degree analysis}
\label{subsec:degreedist}
\begin{figure}[ht]
\centering
\includegraphics[width=0.8\textwidth]{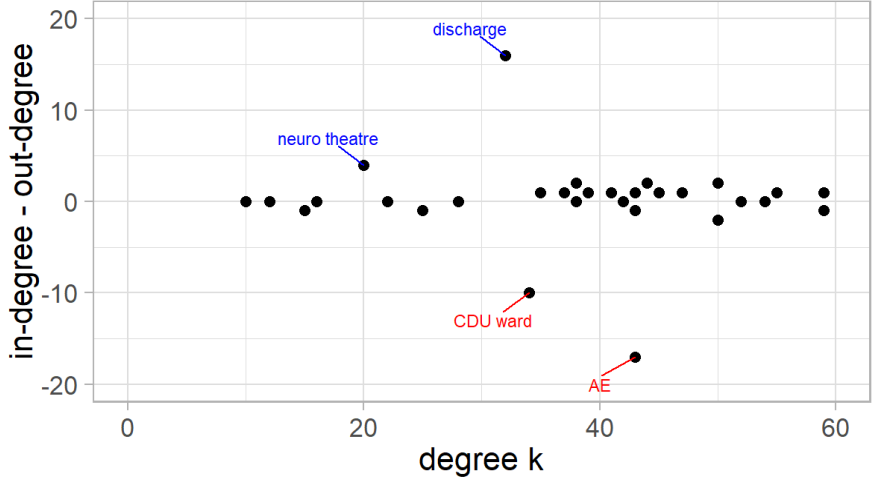}
\caption{\textbf{Net connectivity of nodes}. This shows the unweighted difference in in-degree and out-degree, net connectivity, versus overall degree k for our system of wards. The distribution wards such as A\&E are in the lower half of the graph (red labels) and the receiving wards in the upper (blue labels). Most wards have a balanced traffic. The ward abbreviations are explained in the appendix.}
\label{fig:inoutdegrees}
\end{figure}

The in- and out-degree balance is shown in Figure  \ref{fig:inoutdegrees} for the non-categorised network. Most wards fall on or near the zero line showing that they receive from and send patients to a similar number of places, however a few areas are so called `distributors' (in red) or `receivers' (in blue). The clinical decision unit (CDU ward), a short stay medical ward that receives patients shortly after admission is a  `distributor' as it sends patients to a large variety of locations. The appearance of the neurosurgical theatres as a `receiver' outlier was unexpected. It can be explained by the specialist nature of the neurosurgery service, which means that the patients tend to be transferred to a limited set of wards for post-operative care but due to bed-pressures they may be cared for pre-operatively in a range of settings.

\subsection*{Weights}
\begin{figure}[ht]
\centering
\includegraphics[width=0.8\textwidth]{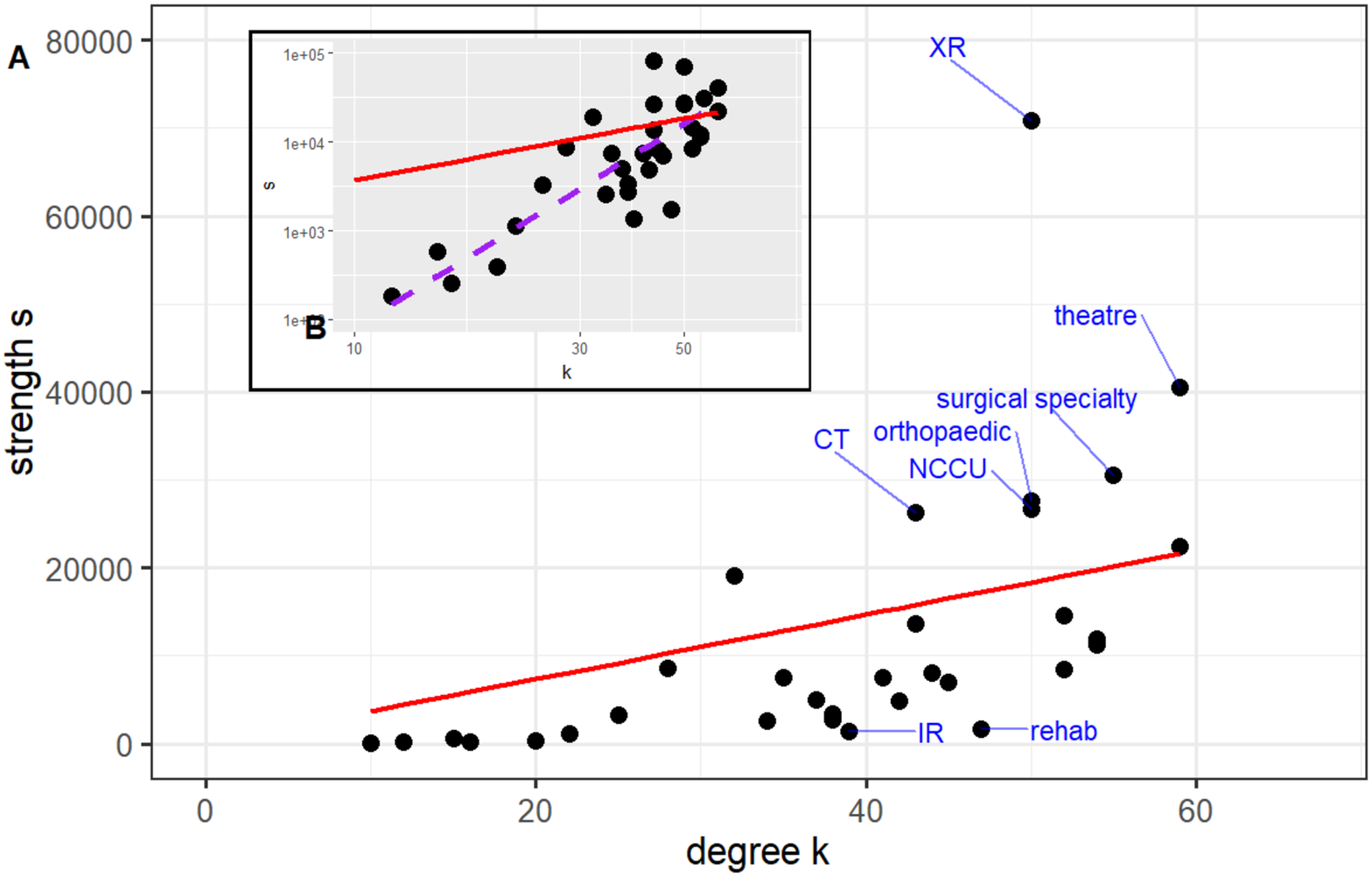}
\caption{\textbf{Relation between traffic and connectivity}. The figure explores the strength (weighted degree) versus degrees for the categorised system of wards. In A: The overall distribution of strength versus degree, where the labelled nodes are outliers with respect to their degree-strength relation, either significantly higher traffic than expected by their connectivity or significantly lower. The graph also shows in red the relation between strength and degree if the weights were uncorrelated. In B: The inlay shows the log-log distribution of strength versus degree with the uncorrelated distribution (red) and the power-law fit (purple dashed) showing that the strength grows faster than the degree. The ward abbreviations are explained in the appendix.}
\label{fig:strengthversusdegree}
\end{figure}

The relationship between the node strength $s$ (weighted degree) and degree in Figure \ref{fig:strengthversusdegree}. Part A of the figure shows the strength versus degrees $k$ for the categorised network. The labelled nodes are selected either due to significantly higher traffic, represented by strength s, than expected by their connectivity (degree) or significantly lower. 

In the absence of correlation between weight and degree,  the strength of a node should be proportional to its degree \cite{Barrat2004} $s \sim k$ (shown by the red line in Figure \ref{fig:strengthversusdegree}), but the data is not well fitted by this distribution, especially at lower degrees. The inlay figure B shows the log-log distribution of strength versus degree again with the uncorrelated line in red and additionally the power-law fit (purple dashed line) for $s \sim k^{\beta}$. The power-law is a much better fit with $\beta \sim 2.1$ showing that the strength of the nodes grows significantly faster than the degree and that higher connectivity wards experience disproportionately more traffic.

\subsection*{Assortativity}
\begin{figure}[ht]
\centering
\includegraphics[width=0.8\textwidth]{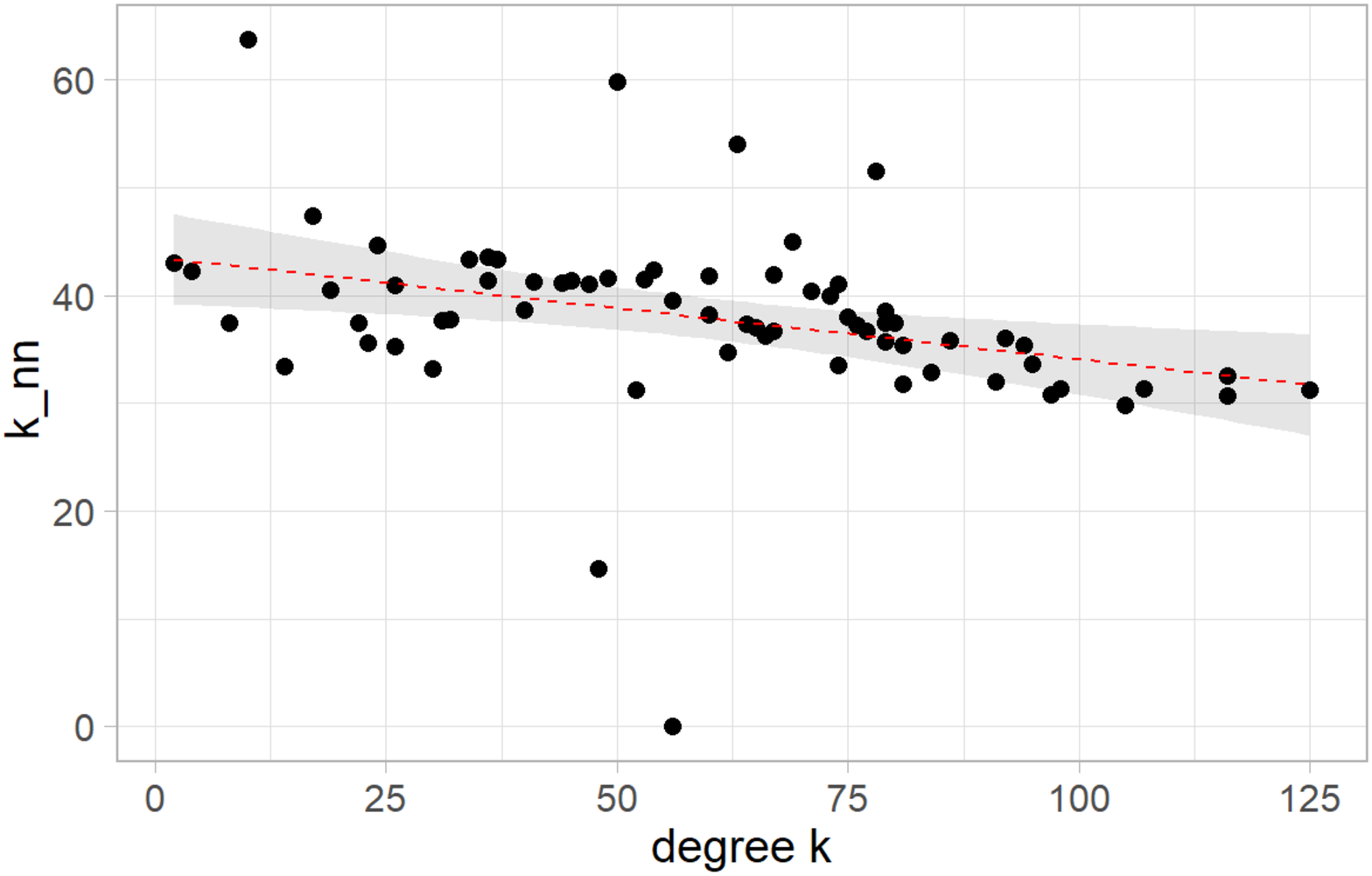}
\caption{\textbf{Assortativity}. The weighted nearest neighbour degree $k_{nn}$ versus degree $k$ for the non-categorised network. It shows the dissociative behaviour of the network where higher degree nodes are connected to lower degree nodes. The best linear fit is overlaid in red.}
\label{fig:assortativity}
\end{figure}

Assortativity $a$, which measures how similar the neighbours of a node are with respect to degree, falls between -1 and 1. We found our system to be dissortative wih assortativity coefficients of $a = -0.20$ and $a=-0.12$ for non-categorised and categorised networks respectively. The dissortative effect is also seen in Figure \ref{fig:assortativity} which shows the average degree of the nearest neighbours, $k_{nn}$, versus degree with the best linear fit overlaid in red. This means that on average high degree nodes are connected to nodes with lower degree and not on average to other highly connected nodes, thus exposing the network to a higher risk of disconnection should one of the highly connected nodes fail \cite{noldus2014}.

\subsection*{Betweeness centrality}
\begin{figure}[ht]
\centering
\includegraphics[width=0.8\textwidth]{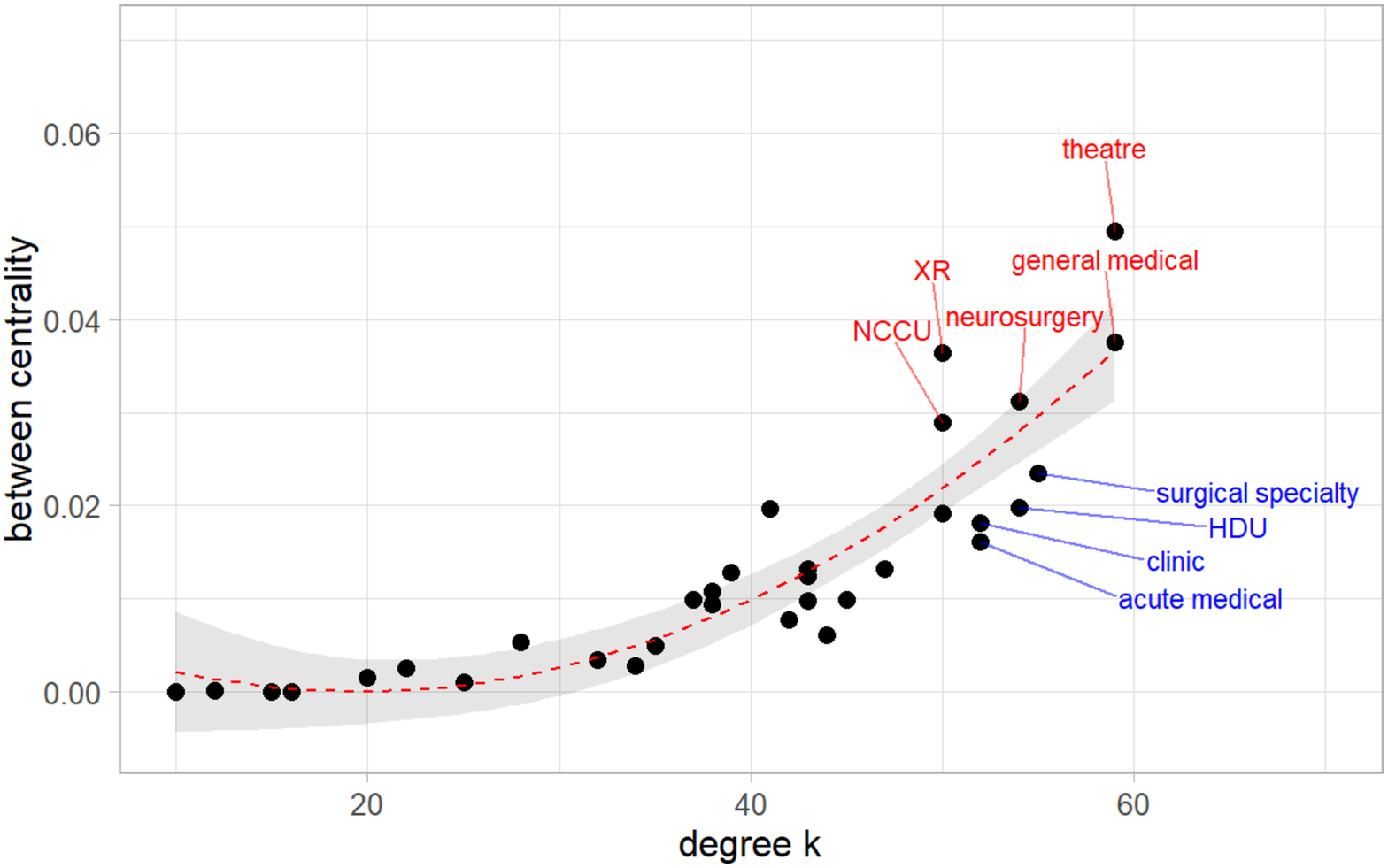}
\caption{\textbf{Betweenness centrality}. The distribution of betweenness centrality versus unweighted degree. The relation is fit by a quadratic equation of the degree - as is commonly seen for randomised networks. The ward abbreviations are explained in the appendix}
\label{fig:betweendegrees}
\end{figure}

Figure \ref{fig:betweendegrees} shows a strong correlation of betweenness centrality with degree. The correlation is shown as a quadratic fit, a feature commonly present in scale-free networks where larger degree nodes have a disproportionately larger betweenness centrality \cite{Mossa2005}. Similarly to above, outliers from this correlation are labelled in red or blue in Figure \ref{fig:betweendegrees} with the group of nodes in red signifying areas that have higher than expected betweenness centrality and are therefore deemed to be essential to the functioning of the system.

We can see that a similar set of nodes is highlighted by the strength-degree relationship (Figure \ref{fig:strengthversusdegree}) and the betweenness centrality one (Figure \ref{fig:betweendegrees}), which is a common occurrence in networks with high reciprocity \cite{Valente2008}. These nodes are well connected, central and experience high traffic. Specifically the examples of radiology (XR), theatres, neuro-critical care (NCCU), general medical and neurosurgical wards show that these ward areas have a disproportionately higher betweenness compared to their connectivity. They are therefore more important to the overall structure of the network than their connectivity may suggest. On the other hand the surgical specialty and acute medical wards have a lower betweenness compared to their connectivity.

We designated the areas with degree in the top 20\% of degrees as hubs and the wards with betweenness centrality within the top 20\% of betweeness centrality as bottlenecks \cite{Yu2007}. This resulted in a set of wards that are both hubs and bottlenecks (hub-bottleneck) - general medical wards, operating theatres, neurosurgical wards - a set that are hubs but not bottlenecks (hub-non-bottleneck) - acute medical wards, clinics, specialty surgical wards and HDU - and non-hub bottlenecks - radiology and NCCU (see Table \ref{tab:hubs}). 

\begin{table}[ht]
\setlength{\tabcolsep}{6.5pt}
\small
\renewcommand{\arraystretch}{1.1}
\centering
\begin{tabular}{|l|l|l|}
\hline
\bf{Hub bottleneck} & \bf{Hub non-bottleneck} & \bf{Non-hub bottleneck} \\
\hline
\hline
surgical specialty wards &  acute medical wards & neuro-critical care \\
\hline
neurosurgical wards & high dependency unit & radiology \\
\hline
general medical wards & clinics/referral appointments & \\
\hline
theatres & & \\
\hline
\end{tabular}
\caption{\textbf{Hubs and bottlenecks}. The hubs and bottlenecks in our system of emergency surgical admissions selected by degree and betweenness centrality.}
\label{tab:hubs}
\end{table}

\subsection*{Scale-free and small world networks}
Apart from analysing the properties of individual nodes we used the above calculations to classify our network as a whole. We used the established parameters $\sigma$ and $\omega$ as described by Telesford et al \cite{Telesford2011} and Humphries and Gurney \cite{Humphries2008}. These two parameters indicate whether a network can be considered to have small-world properties compared to a randomly wired one with similar properties and size. We found that both the original non-categorised network and the categorised network show clear small-world properties: The categorised network had an $\sigma$ of 0.99, where the range of $\sigma$ is from 0 to 1 with values close to 1 representing a small-world network. Also, the value of $\omega$ was 0.0053. Here the range is from -1 to 1 with the values near zero representing small-world networks, close to -1 representing lattice networks and values close to 1 are purely random networks. The non-categorised network similarly had $\sigma$ =0.99 and $\omega$=0.0022, again reflecting its structure as a small-world network.

However due to the high $\gamma$, representing a more rapid drop of nodes with high degrees than other real-world networks \cite{Newman2003}, which tend to have $\gamma$ between 2 and 4, there is a case to be made that this reflects an exponential truncation to a power law \cite{Barabasi1999}. Our system therefore falls into the small-world, scale-free category.

\section*{Discussion}

We demonstrated that our system is a complex and densely connected network. This has important consequences as the connections are crucial to behaviour and therefore resilience. Our values for $\sigma$ and $\omega$ as well as the value of $\gamma>3$ (although this is larger than typical for other real-world networks \cite{Newman2003}) place our system into the category of small world networks \cite{Barabasi2003a}. These are resilient to shocks, retaining this property even when under attack \cite{Davis2003}. However we also found a dissortative structure; common in technological networks \cite{Johnson2010}, implying the presence of a node hierarchy where non-hubs are preferentially connected to hubs \cite{DeMontis2007}. Such structures are highly reliant on their hubs \cite{Newman2003} making these points of vulnerability. The implication for our system is that the removal of a hub from the network would have a significant impact on the workings of the system and is not easily replaced with existing connections.

It is possible to identify hubs and thus potential bottlenecks of our system (Table \ref{tab:hubs}), which is useful in determining where to increase capacity in order to keep the system running smoothly. Combining this information with the areas that have exceptional traffic identified vulnerable areas such as surgical specialty wards, theatres, NCCU and radiology. Failure of an area may be due to closure through infectious outbreak, blockage due to high acuity patients or overload of the area. If these events occur in one of these important nodes the impact on the greater system may be severe and cause problems in unexpected and seemingly unconnected areas. With the connection to the elective surgical service via competition for surgical beds such an event may directly influence the ability of the overall surgical service to function.

The analysis of the traffic between nodes related to connectivity identified areas in the hospital with higher than expected traffic (E.g. orthopaedic and surgical specialty wards and NCCU). This technique may help focus future improvement measures to reduce any delays or improve capacity. The fact that in the whole system well connected wards have disproportionately more traffic is a reflection of the nature of medical services where there is high throughput through certain nodes that are heavily used by different groups and therefore constitute the base of clinical activity, whereas some sub-specialty nodes are only accessed from a small group of nodes for very specific patients. 

In the graphical representation Figure \ref{fig:category_network} we observed the expected importance of the particular nodes such as radiology, the theatre complex and surgical wards and particularly neurosurgery (relatively self-contained in our hospital). Unexpectedly the medical wards, including elderly care also contributed significantly to the functionality of the system. This may reflect both the frequent use of such areas for surgical overflow, incorrect admission triage or the development of surgical problems during the stay.

In Figure \ref{fig:strengthversusdegree}, nodes with $s >> \overline{w} k$ such as theatres or surgical specialty wards represent areas that have high traffic but relatively limited connectivity in contrast to such as IR and rehab which have low traffic compared to their connectivity. This is of interest as it implies that that any impediment of traffic through these high traffic areas will have a strong influence on the other areas associated with them: Their influence on the system is therefore more relevant to individual patients pathways rather than overall behaviour.

Our network representation was developed using retrospective electronic patient records specifically location data which will always contain noise. It is not clear the overall network structure would be preserved under severe shock conditions (E.g. a major incident might severely reduce theatre capacity, an area of vulnerability from our findings). Under such extreme circumstances the overall configuration of the hospital may transiently different from what we found under business-as-usual conditions. However these are rare, and more relevant shocks would be expected to be captured in our model. 

Our model is an aggregate of data over the whole time period and does not describe the dynamic behaviour of the system, for example under conditions of hospital strain. It is possible however to construct time-series of networks \cite{Ebadi2019a} and it may be that such an approach could demonstrate the actual response of the system to external perturbation. Such an approach might also be a useful framework for forecasting critical bed status.

Despite all of these limitations our data appears to show a full picture of the system and give us initial insights in how we can use this approach to inform decisions on service improvement and planning. It would be interesting to compare the configuration of our institution with others so better understand how different hospital configurations might influence potential system resilience. 

\section*{Conclusions}
We were able to use electronic health care records to create a system representation of our service and demonstrated that emergency surgical services are complex systems with scale-free, dissortative small-world network properties. Such networks are robust overall but may be vulnerable to attack at critical hubs. We are able to identify system bottlenecks and this may form the basis to inform service improvement initiatives in a more holistic way. This analysis allowed us to show that new insights into the structure and vulnerabilities of a system of care can be gained by combining network analysis and electronic care records. In the future we hope to extend this work by considering seasonal effects and use the model to better understand the systems behaviour under strain.

\section*{Funding}
KK is supported by a National Institute for Health Research academic clinical fellowship.

\section*{Conflicts of interest}
None declared.

\section*{Data sharing}
The data was collected as part of routine clinical care and the authors cannot make this available openly. Research collaborations will be entertained on application.

\section*{Acknowledgements}
The authors would like to thank Daniel Stubbs for useful discussions and Shaun Hyett for technical assistance in extracting the EHR data. We would also like to thank the Healthcare Design group in the Engineering Department at University of Cambridge for their helpful input.


\section*{Appendix}
\label{sec:suppwards}

\begin{table}[ht]
\setlength{\tabcolsep}{18pt}
\renewcommand{\arraystretch}{1.5}
\centering
\begin{tabular}{|l|l|}
\hline
\bf{Abbreviation} & \bf{Ward area} \\
\hline
\hline
AE or A\&E &  Emergency department \\
\hline
Angio & Angiopgraphy\\
\hline
CDU & clinical decision unit (short stay) \\
\hline
Clinic & Referrals and appointments within the hospital\\
\hline
CT & Computer tomography\\
\hline
Echo & Trans-thoracic echocardiography\\
\hline
HDU & High-dependency unit\\
\hline
ICU & Intensive care \\
\hline
IR & Interventional radiology\\
\hline
MRI & Magnetic Resonance Imaging\\
\hline
NCCU & neuro-critical care \\
\hline
neuro theatre & Neuro-surgical theatres\\
\hline
NP & Neuro-physiology investigations\\
\hline
PET & Positron emission tomography\\
\hline
PICU & Paediatric intensive care \\
\hline
Rehab & Rehabilitation ward\\
\hline
TOE & Trans-oesophageal echocardiography\\
\hline
US & Ultrasound\\
\hline
VA & Vascular access services\\
\hline
XR & Radiology (x-ray department)\\
\hline
\end{tabular}
\caption{Abbreviations used throughout the document for ward areas and services.}
\end{table}

\bibliography{main}

\begin{thebibliography}{10}

\bibitem{NHSEngland2018}
{NHS England and NHS Digital}.
\newblock {Hospital Accident and Emergency Activity}; 2018.

\bibitem{Keogh2018}
Keogh B, Culliford D, Guerrero-Ludue{\~{n}}a R, Monks T.
\newblock {Exploring emergency department 4-hour target performance and
  cancelled elective operations: A regression analysis of routinely collected
  and openly reported NHS trust data}.
\newblock BMJ Open. 2018;.

\bibitem{Wong2018a}
Wong DJN, Harris SK, Moonesinghe SR, Moonesinghe SR, Wong DJN, Harris SK,
  et~al.
\newblock {Cancelled operations: a 7-day cohort study of planned adult
  inpatient surgery in 245 UK National Health Service hospitals}.
\newblock British Journal of Anaesthesia. 2018;121(4):730--738.

\bibitem{Martinez2018}
Martinez DA, Kane EM, Jalalpour M, Scheulen J, Rupani H, Toteja R, et~al.
\newblock {An Electronic Dashboard to Monitor Patient Flow at the Johns Hopkins
  Hospital: Communication of Key Performance Indicators Using the Donabedian
  Model}.
\newblock Journal of Medical Systems. 2018 Aug;42(8):133.

\bibitem{Khanna2013}
Khanna S, Boyle J, Good N, Bugden S, Scott M.
\newblock {Hospital level analysis to improve patient flow.}
\newblock Studies in health technology and informatics. 2013;188:65--71.

\bibitem{Khanna2016}
Khanna S, Sier D, Boyle J, Zeitz K.
\newblock {Discharge timeliness and its impact on hospital crowding and
  emergency department flow performance}.
\newblock Emergency Medicine Australasia. 2016 Apr;28(2):164--170.

\bibitem{Afilal2016}
Afilal M, Yalaoui F, Dugardin F, Amodeo L, Laplanche D, Blua P.
\newblock {Forecasting the Emergency Department Patients Flow}.
\newblock Journal of Medical Systems. 2016 Jul;40(7):175.

\bibitem{Khanna2017}
Khanna S, Boyle J, Good N, Bell A, Lind J.
\newblock {Analysing the emergency department patient journey: Discovery of
  bottlenecks to emergency department patient flow}.
\newblock EMA - Emergency Medicine Australasia. 2017 Feb;29(1):18--23.

\bibitem{Ebadi2019a}
Ebadi A, Tighe PJ, Zhang L, Rashidi P.
\newblock {A quest for the structure of intra- and postoperative surgical team
  networks: does the small-world property evolve over time?}
\newblock Social Network Analysis and Mining. 2019 Dec;9(1):7.

\bibitem{Barabasi2016}
Barab{\'a}si AL, P{\'o}sfai M.
\newblock {Network science}.
\newblock Cambridge University Press; 2016.

\bibitem{Havlin2012}
Havlin S, Kenett DY, Ben-Jacob E, Bunde A, Cohen R, Hermann H, et~al.
\newblock {Challenges in network science: Applications to infrastructures,
  climate, social systems and economics}.
\newblock European Physical Journal: Special Topics. 2012 Nov;214(1):273--293.

\bibitem{Newman2003}
Newman MEJ.
\newblock {The structure and function of complex networks}.
\newblock SIAM Review. 2003;45(2):167--256.

\bibitem{Eubank2004}
Eubank S, Guclu H, {Anil Kumar} VS, Marathe MV, Srinivasan A, Toroczkai Z,
  et~al.
\newblock {Modelling disease outbreaks in realistic urban social networks}.
\newblock Nature. 2004;429(6988):180--184.

\bibitem{Albert1999}
Albert R, Jeong H, Barab{\'a}si AL.
\newblock {Diameter of the World-Wide Web}.
\newblock Nature. 1999;401(6749):130--131.

\bibitem{An2018}
An C, O'Malley AJ, Rockmore DN, Stock CD.
\newblock {Analysis of the U.S. patient referral network}.
\newblock Statistics in Medicine. 2018;37(5):847--866.

\bibitem{Bean2017}
Bean DM, Stringer C, Beeknoo N, Teo J, Dobson RJB.
\newblock {Network analysis of patient flow in two UK acute care hospitals
  identifies key sub-networks for A{\&}E performance}.
\newblock PLOS ONE. 2017 Oct;12(10):e0185912.

\bibitem{Barabasi2003a}
Barab{\'a}si AL, Dezso D, Ravasz E, Yook SH, Oltvai Z.
\newblock {Scale-free and hierarchical structures in complex networks}.
\newblock AIP Conference. 2003;.

\bibitem{Crucitti2004}
Crucitti P, Latora V, Marchiori M, Rapisarda A.
\newblock {Error and attack tolerance of complex networks}.
\newblock In: Physica A: Statistical Mechanics and its Applications. vol. 340.
  North-Holland; 2004. p. 388--394.

\bibitem{Watts1998}
Watts DJ, Strogatz SH.
\newblock {Collective dynamics of ‘small-world' networks}.
\newblock Nature. 1998 Jun;393(6684):440--442.

\bibitem{Humphries2008}
Humphries MD, Gurney K.
\newblock {Network 'small-world-ness': A quantitative method for determining
  canonical network equivalence}.
\newblock PLoS ONE. 2008;3(4).

\bibitem{Telesford2011}
Telesford QK, Joyce KE, Hayasaka S, Burdette JH, Laurienti PJ.
\newblock {The Ubiquity of Small-World Networks}.
\newblock Brain Connectivity. 2011;1(5):367--375.

\bibitem{Cowan2004}
Cowan R, Jonard N.
\newblock {Network structure and the diffusion of knowledge}.
\newblock Journal of Economic Dynamics and Control. 2004 Jun;28(8):1557--1575.

\bibitem{Hagberg2008}
Hagberg AA, Schult DA, Swart PJ.
\newblock {Exploring network structure, dynamics, and function using NetworkX}.
\newblock In: 7th Python in Science Conference (SciPy 2008); 2008. p. 11--15.

\bibitem{Csardi2006}
{Gabor Csardi} TN.
\newblock {The igraph software package for complex network research}.
\newblock InterJournal. 2006;p. 1695.

\bibitem{shannon2003}
Shannon P, Markiel A, Ozier O, Baliga NS, Wang JT, Ramage D, et~al.
\newblock {Cytoscape: A software Environment for integrated models of
  biomolecular interaction networks}.
\newblock Genome Research. 2003 Nov;13(11):2498--2504.

\bibitem{Clauset2009}
Clauset A, Shalizi CR, Newman MEJ.
\newblock {Power-Law Distributions in Empirical Data}.
\newblock SIAM Review. 2009;51(4):661--703.

\bibitem{Efron1979}
Efron B.
\newblock {Bootstrap Methods: Another Look at the Jackknife}.
\newblock The Annals of Statistics. 1979 Jan;7(1):1--26.

\bibitem{powerlaw}
Gillespie C.
\newblock {Title Analysis of Heavy Tailed Distributions}; 2019.

\bibitem{Barrat2004}
Barrat A, Barthelemy M, Pastor-Satorras R, Vespignani A.
\newblock {The architecture of complex weighted networks}.
\newblock Proceedings of the National Academy of Sciences. 2004
  Mar;101(11):3747--3752.

\bibitem{noldus2014}
Noldus R, Mieghem PV.
\newblock {Assortativity in complex networks}.
\newblock Journal of Complex Networks. 2014;3(4):507--542.

\bibitem{Mossa2005}
Mossa S, Turtschi A, Amaral LaN, Guimer{\`{a}} R.
\newblock {The worldwide air transportation network: Anomalous centrality,
  community structure, and cities' global roles.}
\newblock Proceedings of the National Academy of Sciences of the United States
  of America. 2005;102(22):7794--7799.

\bibitem{Valente2008}
Valente TW, Coronges K, Lakon C, Costenbader E.
\newblock {How Correlated Are Network Centrality Measures?}
\newblock Connections (Toronto, Ont). 2008;28(1):16--26.

\bibitem{Yu2007}
Yu H, Kim PM, Sprecher E, Trifonov V, Gerstein M.
\newblock {The importance of bottlenecks in protein networks: Correlation with
  gene essentiality and expression dynamics}.
\newblock PLoS Computational Biology. 2007;3(4):713--720.

\bibitem{Barabasi1999}
Barab{\'a}si AL, Albert R, Jeong H.
\newblock {Mean-field theory for scale-free random networks}.
\newblock Physica A: Statistical Mechanics and its Applications. 1999
  Oct;272(1-2):173--187.

\bibitem{Davis2003}
Davis FD, Yoo M, Baker WE.
\newblock The Small World of the American Corporate Elite, 1982-2001.
\newblock Strategic Organization. 2003 Aug;1(3):301--326.

\bibitem{Johnson2010}
Johnson S, Torres JJ, Marro J, Mu{\~n}oz MA.
\newblock {Entropic origin of disassortativity in complex networks}.
\newblock Physical Review Letters. 2010;104(10).

\bibitem{DeMontis2007}
{De Montis} A, Barth{\'e}lemy M, Chessa A, Vespignani A.
\newblock {The structure of interurban traffic: A weighted network analysis}.
\newblock Environment and Planning B: Planning and Design. 2007;34(5):905--924.

\end{thebibliography}
\bibliographystyle{vancouver}

\end{document}